\documentclass[sigconf]{acmart}
\copyrightyear{2026}
\acmYear{2026}
\setcopyright{cc}
\setcctype{by}
\acmConference[JCDL '26]{The 2026 ACM/IEEE Joint Conference on Digital Libraries}{October 13--16, 2026}{Frisco, TX, USA}
\acmBooktitle{The 2026 ACM/IEEE Joint Conference on Digital Libraries (JCDL '26), October 13--16, 2026, Frisco, TX, USA}
\acmDOI{10.1145/3805696.3846016}
\acmISBN{979-8-4007-2597-5/2026/10}

\usepackage{booktabs}
\usepackage{tikz}
\usetikzlibrary{positioning}
\usepackage{pgfplots}
\pgfplotsset{compat=1.18}
\usepackage{array}
\begin{document}

\title{Same Problem, Different Field: Cross-Domain Solution Import via Domain-Stripped Computational Fingerprints}

\author{Eryk Kulikowski}
\orcid{0000-0002-9967-9732}
\affiliation{%
  \department{LIBIS}
  \institution{KU Leuven}
  \city{Leuven}
  \country{Belgium}}
\email{eryk.kulikowski@kuleuven.be}

\begin{abstract}
The same underlying computational problem is solved across unrelated fields under
different names: recursive Bayesian state estimation appears as a ``Kalman filter'' in control,
``Bayesian forecasting'' in pharmacokinetics, and ``data assimilation'' in geoscience. Topical and
citation-based scientific embeddings cannot see this shared problem. We distill each paper once into
a domain- and method-name-stripped \emph{faceted computational fingerprint}, a free-text mechanism
skeleton plus controlled computational facets. We define a tunable, facet-selectable similarity over it. The goal is
\emph{solution import}: surface cross-field pairs solving the same problem, so a bespoke
implementation can be swapped for another field's standard, specialized solver. On a benchmark of 18
method families across 109 papers, the skeleton lifts cross-domain retrieval average precision over
the abstract from 0.222 to 0.513, and the whole fingerprint reaches 0.557. Strikingly, four trained
scientific embedders all fall below plain abstract+TF-IDF: they encode topical and citation
similarity, the wrong signal for this task. The gain is the representation: the abstract-to-skeleton
swap lifts every embedder, and the pipeline is one cached LLM call per paper plus a cheap embedder. An
interventional re-skin / math-edit test shows the fingerprint tracks the computation, not the field.
On a 501-paper wild corpus, known twins dominate the top of the ranking (23 of the top 30); with planted
pairs excluded from the results, three blind LLM judges rate 3 of the top 5 and 8 of the top 30 pairs genuine
import candidates, and 0 of 30 random ones. The human
verification is the four executed imports: in one, an open standard solver reproduces a bespoke
clinical dosing engine's output. We release the benchmark, the code, and the distillation prompt.
\end{abstract}



\maketitle

\section{Introduction}
\label{sec:intro}
Individualized antibiotic dosing models the body as compartments through which a drug diffuses and
estimates each patient's parameters by a Bayesian procedure. Follow its citations to the underlying
mathematics, and the algorithm estimates a population parameter distribution as a
constrained optimization over a sparse set of support points: the nonparametric maximum-likelihood
estimate of a mixing distribution~\cite{schumitzky}, the same computation used to fit mixtures in machine learning and
astronomy. In its dosing-control form, it resembles LQG~\cite{bayard}, a Kalman filter plus a
linear-quadratic regulator. Neither cited paper says ``mixture model'' or ``Kalman filter''. One
applied problem behaves, depending on the sub-step, like a nonparametric mixture-distribution estimator
and a Kalman filter, yet topical and citation embeddings file all of it under ``pharmacokinetics''. In
Section~\ref{sec:import} we \emph{verify} this import: an open solver for the same
nonparametric-maximum-likelihood problem reproduces the bespoke dosing engine's output on that engine's
own reference data.

This kinship is not an isolated curiosity. A dominant-eigenvector computation on a graph-derived matrix
is ``eigenvector centrality'' in network science, ``PageRank'' on the web, ``Eigenfactor'' in
scientometrics, and a team-strength score in sports analytics. The underlying problem is the same; the vocabulary, the field, and the citation neighborhood are
not.

And the payoff of connecting these re-derivations is documented, not hypothetical: field after field has
recognized its problem as an instance of one already solved elsewhere and imported that field's standard
method to a better result. Epidemic forecasting adopted numerical-weather-prediction data
assimilation~\cite{shaman}; protein contact prediction the inverse Ising/Potts problem from statistical
physics~\cite{ekeberg}; causal panel-data inference matrix completion from recommender
systems~\cite{athey}; radio-interferometric imaging compressed sensing~\cite{wiaux}, outperforming the
standard CLEAN in simulations; crime modeling the seismology aftershock process~\cite{mohler}; and cosmological reconstruction
optimal transport~\cite{frisch}.

Two consequences follow, and both are costly. First, \textbf{implementation quality is uneven}: the
field that owns the canonical method has mature, audited, open-source solvers, while a field that
re-derived it often runs a one-off script or pays for closed software whose core is a decades-old
standard method. Second, and the reason the first persists, the kinship is \textbf{hard for existing
tools to see}: topical and citation embeddings place the two structurally identical papers far apart
(Section~\ref{sec:results}), and the analogy-mining systems built to bridge domains match over the very
surface vocabulary the two papers do not share (Section~\ref{sec:related}). So the shared computation
usually goes unreported and the solver unimported. We address the second problem, which unlocks the
first.

The paper's three terms follow. A paper's \emph{solution} is the computational method it runs to solve
its underlying problem, whether a standard solver or a bespoke implementation. \emph{Solution import}
is replacing the bespoke implementation with another field's standard, specialized solver for the
same underlying problem, the pattern in every example above. An \emph{import candidate} is a
cross-field pair of papers surfaced as solving the same problem, proposed for that swap and awaiting
verification.

\begin{figure}[tbp]
\centering
\begin{tikzpicture}[font=\footnotesize, node distance=3.0mm,
  b/.style={draw, rounded corners, align=center, inner sep=2.5pt, minimum height=6.5mm}]
\node[b](p){paper\\full text};
\node[b, right=of p](d){LLM\\distill};
\node[b, right=of d](s){\textbf{fingerprint}\\skeleton + facets};
\node[b, right=of s](e){cheap\\embed};
\node[b, right=of e](v){retrieve\\then filter};
\draw[->](p)--(d); \draw[->](d)--(s); \draw[->](s)--(e); \draw[->](e)--(v);
\end{tikzpicture}
\caption{The pipeline: one cached LLM call distills each paper into a domain-stripped fingerprint
(mechanism skeleton plus controlled facets); a cheap text-similarity search over the fingerprint text (TF-IDF /
BM25) retrieves candidates, and a facet-agreement filter narrows them to the precise ones.}
\label{fig:pipeline}
\Description{Pipeline diagram: paper full text, to an LLM distillation step, to a fingerprint (mechanism skeleton plus controlled facets), to a cheap embedder, to a retrieve-then-filter step.}
\end{figure}
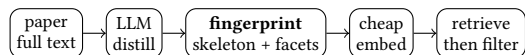

\paragraph{Approach}
We distill each paper once (cached) into a domain-stripped \textbf{faceted computational fingerprint}
(Figure~\ref{fig:pipeline}): a free-text \emph{mechanism skeleton} that answers ``what does this compute, and how'' with the domain
vocabulary and canonical method names removed. The fingerprint adds a small set of \emph{controlled computational
facets} (computational structure or motif, data object, inference, problem form, outcome
\looseness=-1 distribution, complexity). A tunable, facet-selectable \textbf{similarity} over the fingerprint is the
operator, and it needs no special infrastructure: a standard BM25/TF-IDF search over the fingerprint text
retrieves candidate twins at high recall, the facet conjunction filters them for precision, and a human
confirms the survivors.

\paragraph{Contributions}
The primitives are deliberately standard (an LLM distillation pass, off-the-shelf embedders, an
exact-match filter); the claims are the domain-stripped representation, the validated similarity over it,
and the benchmark. Section~\ref{sec:levers} separates each from prior art.
\begin{enumerate}
\item \looseness=-1 \textbf{The faceted computational fingerprint and a facet-selectable similarity over it}
(Section~\ref{sec:method}), with evidence that \emph{the representation carries the result}: a
domain-stripped fingerprint surfaces same-problem, different-field pairs that SPECTER, SPECTER2, SciNCL,
and SemCSE systematically miss. An interventional re-skin / math-edit test shows the fingerprint keys on the
computation, not the field (Section~\ref{sec:perturb}). The gain is neither the labeling (facet
classification on its own is the weakest of our signals, Section~\ref{sec:faceted}) nor an engineered
similarity (a tunable precision/recall operator, not a gain over the plain fingerprint).
\item \textbf{A retrieve-then-filter operator}: the fingerprint text retrieves broadly (the high-recall stage, though recall is bounded) and the
facet conjunction lifts precision several-fold on the retrieved pool, a gain measured on the curated
benchmark (Table~\ref{tab:frontier}). One substrate thus supports many
operators tuned along an explicit precision/recall frontier (Sections~\ref{sec:method},
\ref{sec:faceted}).
\item \textbf{A released cross-domain solution-import benchmark and an open methodology}: 18 method
families across 109 papers and a 501-paper noisy corpus, with topical distractors and name-free
exemplars whose papers never name the method, checked in the wild (known twins fill 23 of the top 30; with planted pairs excluded, three blind LLM judges
rate 3 of the top 5 and 8 of the top 30 genuine, 0 of 30 random; Section~\ref{sec:modeb}).
Plus the finding that for this operator a cheap, coarse-consistent distiller beats a frontier one, because
consistent labels collide where precise ones split (Sections~\ref{sec:benchmark}, \ref{sec:method-results}).
\item \textbf{Four executed cross-domain imports} (Section~\ref{sec:import}): a worked clinical case
(a bespoke dosing estimator that computes a nonparametric mixture / support-point estimate, with an
executed open-solver reproduction), plus three more across distinct computational cores, each
reproducible offline and surfaced by the fingerprint, framed candidate-not-proof.
\end{enumerate}

A scope note we carry throughout: the curated benchmark is built from cores already known to be swappable,
so it measures retrieval \emph{given} that a cross-field twin exists. Whether an arbitrary paper has one
is the harder open-world question, which the in-the-wild study (Section~\ref{sec:modeb}) probes directly.

\section{Related work}
\label{sec:related}
Three lines of prior work touch the problem: systems that bridge domains (analogy mining,
literature-based discovery), representations of scientific documents (embeddings, facets, method
extraction), and retrieval that lets the user choose what counts as similar. All of them keep the
domain vocabulary and the method names in the representation and match on topic, theme, or purpose.
We strip both and match on the computation, so the topical embeddings below are our baselines rather
than our method. Matching on bare structure is established wherever a formal artifact exists to
match: mathematical information retrieval indexes expressions by their symbolic
structure~\cite{mathir}, and clone detection finds code fragments that perform the same computation
under different identifiers~\cite{clone}. A paper's prose supplies no such artifact; recovering the
computation from the prose is our move.

\paragraph{Bridging domains: analogy and literature-based discovery}
Gentner's structure-mapping~\cite{gentner} and its retrieval engines SME and MAC/FAC~\cite{macfac}
are the classical antecedents of matching on structure rather than surface. Analogy-mining systems
apply the idea to papers by decomposing them into purpose and mechanism aspects over their surface
text: Hope and Shahaf~\cite{hope}, SOLVENT~\cite{solvent}, the cross-knowledge-domain retrieval of
Kang et al.~\cite{kang}, and the Analogy Search Engine~\cite{ase}. They retain the domain vocabulary
and match on purpose (same purpose, different field); we strip the vocabulary and the method name and
match on the computation. Swanson's literature-based discovery is the other lineage: logically
connected knowledge can lie undiscovered across non-interacting
literatures~\cite{swansonupk,swanson}, operationalized as shared intermediate B-terms in
ARROWSMITH~\cite{swansonsmalheiser} and matured since with standard models and rediscovery-style
evaluations~\cite{henrymcinnes,yetisgen}. LBD links two literatures through a path of shared
concepts; the pairs we target share no terms, no citations, and no concepts, a connection that term
overlap cannot make. The phenomenon itself is Merton's multiple discovery~\cite{merton}. Automated
detectors of simultaneous discovery key on shared citation context~\cite{kimahn} and so surface only
same-topic twins; the cross-field computational case has, to our knowledge, been documented only by
hand-curated survey~\cite{convergent}.

\paragraph{Scientific document embeddings}
SPECTER~\cite{specter}, SPECTER2~\cite{specter2}, and SciNCL~\cite{scincl} embed title and abstract
by citation-based contrastive learning. They capture topical similarity, and their training signal,
the citation neighborhood, separates fields by construction; they are our baselines.
SemCSE-Multi~\cite{semcse} is the closest in mechanism, LLM-written aspect-specific summaries
embedded contrastively by a fine-tuned encoder, and one might expect it to subsume our skeleton as a
method aspect. It enriches and names \emph{within} a domain, the opposite of our move, never uses
inter-aspect disagreement (same computation, different field) as a signal, and ships only
domain-bound checkpoints, so we run its base encoder as the closest public proxy
(Section~\ref{sec:semcse}). Boyack and Klavans~\cite{boyack} show on 9 million documents that text
and citation clusterings diverge and are complementary; computational structure is a third axis
both miss.

\paragraph{Faceted representations and method extraction}\looseness=-1
Aspect-based document similarity~\cite{ostendorff} scores pairs over a fixed aspect set (task,
method, dataset); Aspire~\cite{aspire} learns one vector per facet; QA-Emb~\cite{qaemb} and
CQG-MBQA~\cite{cqgmbqa} make each dimension an LLM's yes/no answer to a shared question bank; and
recent faceted scientific retrieval decomposes papers into facets at indexing time, blended for
query-by-example~\cite{fable} or reweighted into a single vector~\cite{flew}. A parallel line
extracts per-paper method tuples for knowledge graphs, SciERC~\cite{scierc} and SciREX~\cite{scirex},
and methodology-inspiration retrieval~\cite{mir} ranks prior methods for a new problem. These facets
are topical or task-oriented, largely predefined, and applied within a field, and the retrieval
variants learn from the citation graph, which cannot reach independently re-invented twins that never
cite each other. Ours are domain-stripped \emph{computational} facets, distilled rather than
predefined, with a similarity that selects among them.

\paragraph{LLM-generated representations and idea recombination}
That a generated artifact rather than the embedder can carry the signal underlies HyDE~\cite{hyde},
which embeds an LLM-written hypothetical document for a query; we distill a persistent, domain- and
method-name-stripped representation of each corpus document, once. Scideator~\cite{scideator} has an
LLM distill purpose-mechanism facets for human ideation, the nearest occupant of ``an LLM distills
facets once'', though its facets are rhetorical, unvectorized, and unevaluated for retrieval.
CHIMERA~\cite{chimera} mines \emph{author-stated} recombinations by supervised extraction;
SciMuse~\cite{scimuse} generates cross-domain ideas from concept co-occurrence, SciMON~\cite{scimon}
optimizes generated ideas for novelty, and concurrent work reformulates a research goal into
domain-agnostic conceptual problems~\cite{ideacatalyst}; novelty-of-science work scores atypical
citation combinations~\cite{uzzi} or embedding-distance novelty~\cite{shibayama}. All of these
recombine or score from what authors state or cite. We retrieve isomorphisms the two papers never
state: that a political-science ideal-point estimator and a recommender-systems matrix factorization
recover the same low-dimensional embedding is implicit in the ideal-point model's own item-response
formulation~\cite{clinton}, yet neither paper names, cites, or shares vocabulary with the other, so
no extractor of author-stated links can reach the pair.

\paragraph{Conditional similarity and retrieve-then-filter}
Several systems let the user tune what counts as similar: Conditional Similarity
Networks~\cite{csn} select a learned masked subspace, conditional semantic textual
similarity~\cite{csts} conditions on a free-text aspect, and embedding-based retrieval applies a
score threshold~\cite{relfilter}. Selecting and counting named facets is ordinary faceted search.
What differs here is what the facets \emph{are}: auto-distilled descriptors of the computation with
the domain and the method name removed, so the knob turns on structure and the precision/recall
frontier it traces runs across fields rather than within one.

\paragraph{Positioning}
The ingredient that makes the representation feasible now is the LLM: re-describing a paper's
computation in domain-neutral terms, without method names, is a comprehension-and-paraphrase task
that became reliable at corpus scale only with instruction-tuned models. The contribution is not the
model but the representation it enables and the task it makes measurable, which is why a
bag-of-words over the resulting skeletons already beats trained scientific embedders over abstracts.

\section{The faceted computational fingerprint}
\label{sec:method}
The approach in one line: distill each paper once into a domain-stripped fingerprint, then define a
tunable, facet-selectable similarity over it. The fingerprint has two parts from one cached LLM call: a
free-text \emph{mechanism skeleton} (the bulk of the retrieval signal) and a set of \emph{controlled
computational facets} (the high-precision filter). A separate verifiability facet set is distilled
too, but it feeds a value metric rather than the similarity (Section~\ref{sec:future}).

\subsection{The mechanism skeleton (the retrieval signal)}
\label{sec:skeleton}
The skeleton is an LLM re-description of a paper that answers ``what does this compute, and how'',
stripped of (a) domain vocabulary and (b) the canonical names of any methods used, produced once per
paper and cached. The shipped fingerprints are distilled by Claude Haiku, the cheap hosted tier;
Section~\ref{sec:method-results} swaps it for a frontier model (Opus) and a local open-weights model
(qwen3:14b) and measures what the choice costs; full model-independence is not claimed. The operative instruction, abridged here (the full
prompt ships with the code), is:

\begin{quote}\small\itshape
Describe what this paper computes: the objects operated on, the quantity estimated or optimized, and
the mathematical structure of the procedure. Use domain-neutral language: do not name the application
field, and do not name any canonical method (no ``Kalman filter'', ``SVM'', ``PageRank''). If a named
method is used, describe its mathematics instead: PageRank, for instance, becomes ``the stationary
distribution of a damped random walk on a weighted graph''.
\end{quote}

This has a property crucial for evaluation: for the skeletons that the audit finds fully name-free (Section~\ref{sec:audit}), a
cross-domain match cannot be the embedder (or a reader) recognizing a famous name; it can only be
abstraction to shared structure.
Figure~\ref{fig:skeleton-example} shows the transformation on the dosing paper of Section~\ref{sec:intro}:
the shipped skeleton retains no drug, patient, or dose vocabulary and never says ``NPAG'' or ``EM''; it
describes a mixing distribution over latent per-entity parameters whose maximum-likelihood estimate is
discrete on at most as many support points as entities, the mathematics that makes the paper
computationally isomorphic to nonparametric mixture estimation in machine learning and astronomy.

\begin{figure}[tbp]
\small
\textbf{Abstract (topic):} two nonparametric methods for population pharmacokinetic modeling,
compared on simulated and clinical data.

\medskip
\textbf{Skeleton (excerpt, shipped fingerprint):} \emph{``The method estimates an unknown mixing distribution over latent per-entity parameter vectors in a hierarchical model. Each entity yields a vector of observations equal to a known nonlinear function of its latent parameter vector plus additive zero-mean Gaussian noise whose covariance may depend on those parameters, and the latent vectors are independent draws from a common unknown distribution $F$. The marginal log-likelihood is a sum over entities of the logarithm of an integral of each entity's conditional density against $F$.''}
\caption{Distilling the dosing paper. The abstract is about population pharmacokinetics; the mechanism is
nonparametric mixture estimation over support points, with neither the field nor the method named.}
\label{fig:skeleton-example}
\Description{A population-pharmacokinetics paper's abstract beside its domain-stripped mechanism skeleton, which describes nonparametric maximum-likelihood estimation of a mixing distribution without naming the field or the method.}
\end{figure}

\subsection{The controlled computational facets (the filter layer)}
\label{sec:facets}
On top of the skeleton, the same call emits a labeled value for six controlled computational facets plus a DOMAIN
field axis (seven in all), each a domain-neutral descriptor written for embedding. Each is listed with its grounding taxonomy and
whether it drives precision (P) or recall (R): \textbf{DOMAIN} (R; AMS MSC2020, coarsened) is the
topical axis a plain abstract embedding already captures, included as the recall floor;
\textbf{STRUCTURE/MOTIF} (P, the centerpiece) is grounded in the Berkeley ``13 dwarfs'' of
computation~\cite{dwarfs}, a published taxonomy of computational \emph{patterns} at a domain-independent
abstraction, and carries the core hypothesis that shared computational structure, not shared domain, marks an import candidate; \textbf{DATA\_OBJECT} (P) is the primary structure operated on (dense
or sparse matrix, grid, mesh, graph, point set, sequence, tree, set, field); \textbf{INFERENCE}
(mixed) is how uncertainty is handled; \textbf{PROBLEM\_FORM} (R, the orthogonalizer) is the abstract
goal (estimation, prediction, optimization, search, simulation, control); \textbf{DISTRIBUTION} (mixed)
carries an internal disagreement, a measured side and an assumed side, the gap being a second operator;
and \textbf{COMPLEXITY} (P, sparse) records a stated guarantee when present. STRUCTURE and DISTRIBUTION
are the precision-drivers; DOMAIN and PROBLEM\_FORM are the recall-drivers, the split the design
predicted, which the per-facet weights confirm (Section~\ref{sec:faceted}). The facets are not seven
independent axes, and we do not assume orthogonality; the per-facet design is
itself a robustness argument, since a corrupted facet does not poison the others as a single fused
vector would. The controlled vocabulary is not hand-designed but bootstrapped from the model's own
free-text categorization and curated (Section~\ref{sec:method-results}).

\subsection{The tunable similarity and retrieve-then-filter}
\label{sec:distance}
The similarity combines the fingerprint's two halves. The text half is a cosine over the whole fingerprint text (the high-recall retriever). The facet half scores how many of the controlled facets the two papers agree on,
where a facet ``agrees'' when both carry the same controlled value (an exact-match Hamming count). The released form
behind Table~\ref{tab:frontier} tests the two values by MiniLM cosine $\ge 0.85$; exact matching reproduces it to within 0.015 and runs far faster.
We compare three ways to aggregate the per-facet agreements into that count, a raw
equal-weight count (the parameter-free baseline), a per-facet log-likelihood-ratio weighting learned
in-sample from the labels (log-LR), and a Boolean \emph{conjunction} (require agreement on at least $k$ facets of a
chosen set; the count $k$ is the operator's one knob, instantiated in Section~\ref{sec:faceted} as $k$
from 0 to 4 over the four core facets); the conjunction is the one the operator
uses, the close-on-computation, far-on-field rule, with the other two as baselines. The fingerprint cosine
and the facet agreement are then combined in one of two ways. The first is a single blended similarity,
\[
\mathrm{sim} = \alpha\,z(\text{fingerprint cosine}) + (1-\alpha)\,z(\text{facet agreement}),
\]
each half $z$-normalized in-sample (no held-out split), with the best mixing weight near $\alpha=0.4$.

\looseness=-1 The second is a \emph{retrieve-then-filter}, and since it is the operator we state it precisely. Let
$C_N$ be the $N$ highest-scoring cross-field pairs $(q,p)$ by fingerprint-text similarity, the
high-recall pool ($N{=}1000$ throughout Section~\ref{sec:faceted}). For a controlled facet $f$, let
$\mathrm{agree}_f(q,p)\in\{0,1\}$ indicate an exact match of the controlled value for $f$ (a missing value never agrees).
Given the chosen facet set $F$ (the four core facets in our instantiation) and the count $k$, the
operator returns
\[
\{\,(q,p) \in C_N \;:\; \textstyle\sum_{f\in F}\mathrm{agree}_f(q,p)\;\ge\;k\,\},
\]
so raising $k$ trades recall for precision along the frontier of Section~\ref{sec:faceted}. Operationally the operator needs no special infrastructure: per query, the retrieve stage is a standard BM25 query
over the fingerprint text, and the facet conjunction is a keyword-field filter over the controlled facets
(a \texttt{minimum\_should\_match} of $k$ over the facet terms) applied as a second step to the top-ranked BM25 hits on the
Elasticsearch/Lucene stack a repository already operates. The one addition is the cached per-paper distillation, and it is the pipeline's whole
LLM cost: one distillation per paper, batched at about a dozen papers per call (roughly forty calls for the
501-paper corpus), emitting about a hundred words of mechanism prose plus seven short facet values.
Retrieval, the facet filter, and the clustering and rule-mining
views all reuse the same cached fingerprint, so re-distillation is only ever a schema or model upgrade. The
best distiller here is also the cheapest hosted tier (Section~\ref{sec:method-results}), and the
local qwen3:14b arm runs on a single consumer GPU, so the stage is affordable at library scale in
either hosted or self-hosted form; the embedding and retrieval stages are free, local, and standard.

\subsection{Two levers, and what is new}
\label{sec:levers}
The system has two knobs on different axes: the distillation \textbf{prompt} sets how good the vectors
are and lifts performance everywhere, while the facet \textbf{subset} and aggregation pick the operating
point and move along the precision/recall trade-off. The best configuration therefore depends on the
operator one wants; the computational-isomorphism operator (high structure similarity, far in domain) is
the one we validate here. On novelty, we separate what is new from what is not. Decomposing a paper into facets and learning one vector per facet are
established~\cite{aspire,ostendorff}; having an LLM distill the facets in one pass is recent
(Scideator~\cite{scideator}, Section~\ref{sec:related}). Choosing which facet must agree is itself established, as
aspect-conditioned retrieval~\cite{aspire} and, in the analogy line, as a near-on-purpose,
far-on-domain rule~\cite{hope,ase}. What we do not locate in prior art is the conjunction: a selectable
\emph{subset} of \emph{domain-stripped computational} facets whose choice traces an explicit
precision/recall frontier, aimed at cross-field \emph{solution import}.

\section{The benchmark and data}
\label{sec:benchmark}
We report on two corpora with different jobs and never conflate them. The \textbf{curated retrieval
benchmark} is 109 papers across 18 method families with 92 queries (every member paper, each with a twin in the
closed candidate pool, so its labels are complete and its P@1 / P@5 / MRR / AUROC / AP / ARI are clean); it is where we
\emph{compare methods}. Our headline metric is AP (average precision): it measures how cleanly the true
cross-domain twin pairs concentrate at the top of the ranked list, where 1.0 means every twin ranks
above every non-twin and about 0.04 is random at this prevalence; we keep AUROC alongside it for
continuity, but do not lead with it, since a high AUROC can coexist with a poor top of the ranked list,
which is the part a retrieval user actually sees. The
\textbf{extended corpus} is 501 papers (109 labeled seeds and 392 unlabeled in-the-wild papers, of which 64 are deliberately
non-mathematical); it is
the \emph{wild run}. The operator skips as a candidate every paper whose STRUCTURE facet is ``none'' (80 of the
501, including one benchmark member and 40 of the 64 non-mathematical papers) and ranks the remaining 82{,}786 cross-field pairs once. Only the families are
labeled there (206 of the 210 twin pairs remain), so we read that ranking two ways: \emph{with planted pairs
included}, detection of known twins among unrelated papers; \emph{with planted pairs excluded from the
results} (every pair of two benchmark papers set aside), discovery on unlabeled pairs. Precision and AP here
are not a benchmark.

\paragraph{Families and distant domains}
Each family (Table~\ref{tab:families}) is one mathematical core instantiated across multiple fields, and
chosen for a specific reason: each is a core we know is \emph{swappable}, with a standard, specialized
implementation available. So when the fingerprint pairs a paper's bespoke version with the standard one
in another field, that pair is an actionable \emph{import} candidate, not merely a curiosity, and the
families serve as ground truth precisely because the swap is already known to work.
Two families deliberately reach into statistics-using fields far from the core's home: political
methodology (ideal-point estimation as a low-dimensional latent embedding of a vote matrix, in the PCA/SVD family)
and digital-humanities stylometry (a linear text classifier). If the fingerprint matches these to
their structural twins in genetics, recommender systems, or NLP, it is doing structural abstraction,
not topical clustering.

\begin{table}[tbp]
\caption{The 18 method families and example fields. Each family is one mathematical core re-derived
across unrelated domains; the per-family paper counts ($n$, members plus their distractors) sum to 109.}
\label{tab:families}
\small
\begin{tabular}{>{\raggedright\arraybackslash}p{3.0cm}c>{\raggedright\arraybackslash}p{3.3cm}}
\toprule
Family (mathematical core) & $n$ & Example fields \\
\midrule
recursive Bayes / Kalman & 10 & control, finance, pharmacokinetics, aerospace, epidemiology \\
EM / latent mixture & 7 & ML, astronomy, pharmacokinetics \\
eigenvector centrality & 8 & web, scientometrics, neuroscience, sports \\
Gaussian process & 8 & ML, geostatistics, materials, exoplanets \\
Fourier / spectral & 6 & astronomy, econometrics, gravitational waves \\
IRT / logistic / Bradley-Terry & 6 & education, psychometrics, sports \\
linear text classifier & 7 & NLP, digital humanities, genomics, clinical \\
low-rank PCA / SVD & 8 & genetics, recsys, political methodology, causal inference \\
sparse / L1 / LASSO & 7 & finance, genomics, MRI, radio astronomy \\
dynamic programming / Viterbi & 7 & bioinformatics, communications, speech \\
diffusion / heat & 6 & physics, fluids, oncology, anthropology \\
HMM & 6 & genomics, ecology, econometrics, structural biology \\
preferential attachment & 6 & networks, economics, linguistics, cities \\
replicator / Lotka-Volterra & 6 & ecology, economics, finance, sociolinguistics \\
MCMC / Metropolis & 5 & cosmology, physics, cryptography \\
inverse Ising / Potts & 2 & structural biology, neuroscience \\
self-exciting / Hawkes & 2 & criminology, seismology \\
optimal transport & 2 & cosmology, machine learning \\
\bottomrule
\end{tabular}
\end{table}

\paragraph{Distractors and name-free exemplars}
Each family includes topical \emph{distractors}, papers in the same field as a member but using
different mathematics (hard negatives a topical embedder places near the member). Where findable, we
include papers that use a family's method but never name it; the pharmacokinetics trace of Section~\ref{sec:intro} supplies a gold exemplar (an unnamed
nonparametric-MLE / support-point estimator), as does kriging as an unnamed
Gaussian process.

\paragraph{Labeling, task, and metrics}
A positive (twin) is a \emph{swappable} pair: two papers attacking the same underlying problem closely
enough that one could import the other's solution, a standard specialized solver in place of a bespoke
one, and compare results (the import scenario, demonstrated later with vancomycin dosing). Labels are
hand-curated at the family level (each paper
assigned one core, field, and role); each assignment is a verifiable determination from the paper's
mathematics, released as the skeleton, not a subjective judgment, so the labels rest on verifiability
rather than annotator consensus. Twin \emph{pairs} are then \emph{derived}, not chosen: every
cross-field pair (assigned fields differ; same-field pairs are never ranked) sharing a family label is a twin, which yields 210 of the 5{,}812 cross-field pairs
over the 109 papers, and because the pool is closed and every paper carries a label, the pair labels
are complete. Curated numbers keep every paper and
every pair, including the ten pairs among the five no-method distractors that the wild run's candidate skip
would remove; dropping only those ten pairs would move the fingerprint's AP from 0.557 to 0.595 and leave the
skeleton's and the abstract's unchanged, so the reported figures are the conservative ones. On
the 501-paper corpus only the planted twins are labeled, so a label-based precision or AP would count every
unlabeled true pair as an error (Section~\ref{sec:modeb} finds such pairs at the top); there the one scale
number that stays honest is recall of the \emph{known} twins. The corpus is defined by versioned link-lists and
reproduces from open sources.

\section{Results: the retrieve layer}
\label{sec:results}
We first establish the skeleton as a cross-domain retriever that beats topical and citation embedders,
then turn to the faceted operator (Section~\ref{sec:faceted}).

\subsection{Retrieval on the curated benchmark}
\label{sec:semcse}
On the curated benchmark the representation effect is clear under every embedder (Table~\ref{tab:grid}):
the skeleton beats the abstract under all eight, the whole fingerprint is at or above the skeleton
everywhere except MiniLM and SPECTER2, and our cheap whole fingerprint + TF-IDF (AP 0.557) tops the
entire grid. The thesis is one row of it: the dedicated scientific embedders all fall below
plain abstract+TF-IDF (0.222) on abstracts, SPECTER lowest at 0.095,
because their citation-based training clusters papers by field, the opposite of what cross-domain twins
need; even strong general-purpose embedders (Qwen3-Embedding~\cite{qwen3emb}, E5-large-v2) do no better
than the cheap baseline on abstracts, within noise. Each baseline is run locally on the full benchmark in its best configuration (SPECTER2's
proximity adapter, E5's task-query prefix), so the gap is not a handicap we imposed. The same
domain-stripping then \emph{rescues} them, SPECTER more than tripling to 0.334 on the whole fingerprint.
The closest recent prior art,
SemCSE-Multi~\cite{semcse}, ships only domain-specific checkpoints, so its base encoder~\cite{semcsebase}
is the fair stand-in, and it loses too (abstract 0.141).

A paired bootstrap over the 109 papers ($B{=}2000$, fixed representation) confirms the ordering is
significant: the whole fingerprint+TF-IDF reaches AP 0.557 (95\% CI $[0.407, 0.720]$), and its paired
margin over every baseline excludes zero, from $+0.33$ over Qwen3-Embedding and $+0.34$ over abstract+TF-IDF
to $+0.46$ over SPECTER (all $p<0.001$, all surviving Holm--Bonferroni). The increment over the bare mechanism
skeleton ($+0.044\,[-0.081,+0.156]$, $p=0.38$) is positive but not significant at this sample size, so the
skeleton carries most of the signal and the facets add a smaller margin. Because the 210 twins come from
only 18 independent cores, we also resample at the \emph{family} level (cluster bootstrap over the cores):
the headline survives, whole fingerprint over abstract $+0.335\,[+0.230,+0.527]$ ($p{<}0.001$), so the
effect is not an artifact of treating papers as independent. Nor is it topical leakage from the
recall-floor DOMAIN facet: dropping DOMAIN from the fingerprint still gives AP 0.522 (above the bare
skeleton), while DOMAIN alone scores only 0.094, so the win is the domain-stripped mechanism and
computational facets, not the field label.

\begin{table*}[tbp]
\caption{AP on the curated benchmark (109 papers, 5{,}812 cross-field pairs, 210 twins). The skeleton beats
the abstract under every embedder, and the whole fingerprint+TF-IDF (0.557) tops the grid; the four trained
scientific embedders all fall below plain abstract+TF-IDF, yet the fingerprint lifts all four
(SPECTER 0.095 to 0.334).}
\label{tab:grid}
\small
\begin{tabular}{lcccccccc}
\toprule
Representation & TF-IDF & MiniLM & SPECTER & SPECTER2 & SciNCL & SemCSE & Qwen3-Emb & E5-large \\
\midrule
abstract & 0.222 & 0.203 & 0.095 & 0.182 & 0.149 & 0.141 & 0.226 & 0.144 \\
skeleton & 0.513 & 0.407 & 0.300 & 0.347 & 0.332 & 0.310 & 0.439 & 0.422 \\
whole fingerprint (ours) & \textbf{0.557} & 0.384 & 0.334 & 0.319 & 0.337 & 0.337 & 0.462 & 0.450 \\
\bottomrule
\end{tabular}
\end{table*}

\subsection{Controlled comparison: same embedder, different input}
The cleanest controlled demonstration fixes the embedder and swaps only the input:
under a plain bag-of-words the cross-domain twins go from AP 0.222 to 0.513 (and P@1 0.55 to 0.82, MRR
0.67 to 0.86): a near-useless cross-domain ranker becomes a strong one purely by
changing what it reads. The gaps are significant (paired bootstrap over
queries, $B{=}2000$, 92 queries): swapping the abstract for the skeleton lifts MRR by
$+0.194\,[+0.109,+0.279]$ and P@1 by $+0.261\,[+0.152,+0.370]$ under TF-IDF. Every interval excludes zero.

\subsection{Ablation, distiller robustness, and scale}
\label{sec:ablation}
Two skeleton arms, both computational-core and method-name-free, differ only in whether domain
vocabulary is stripped (Table~\ref{tab:ablation}, full 109-paper benchmark, the locally-reproducible
qwen3 distiller so the two arms differ only in the strip/keep instruction). Re-describing the paper around
its computation is the main lift (abstract to keep-domain skeleton); stripping the domain vocabulary then
adds a margin that grows with the embedder's topicality (monotone across the three ablated): small under a lexical bag-of-words
($+0.038\,[-0.096,+0.164]$, TF-IDF), larger under MiniLM ($+0.095$), and significant under the
citation-trained SPECTER ($+0.104\,[+0.002,+0.237]$), which otherwise re-injects field similarity from
the retained domain words. Domain-stripping is thus what lets the representation resist topical embedders;
under a lexical embedder its marginal effect is small and the re-description carries most of the lift.
The abstract-to-skeleton retrieval gain is not an artifact of one distiller: it holds for both the Haiku fingerprints
(Table~\ref{tab:grid}) and the locally-reproducible qwen3 distiller (Table~\ref{tab:ablation}), where the
skeleton beats the abstract under every embedder; it is the \emph{faceted operator}, not the skeleton
retrieval, that is distiller-sensitive (Section~\ref{sec:method-results}). The headline is also robust to benchmark composition: the
three two-member families (inverse Ising/Potts, Hawkes, optimal transport) each contribute a single
cross-field pair, and excluding all three leaves AP essentially unchanged (0.568 vs 0.557).
Finally, we apply the similarity \emph{in the wild} (detection, not a scored benchmark; Section~\ref{sec:modeb} judges what
it surfaces). Two signals survive at scale. First, the known
swappable twins still surface near the top: recall of the labeled twins is 0.619 in the top 1000 of the
82{,}786 ranked pairs, against 0.229 for the abstract. Second, the STRUCTURE-``none'' skip catches 40 of the 64
deliberately non-mathematical papers; none of the 24 that pass reaches the top 30 of either reading.

\begin{table}[tbp]
\caption{Domain-stripping ablation (AP, full curated benchmark, qwen3 distiller; keep-domain vs stripped
skeletons differ only in whether domain words are retained). The gain widens as the embedder grows more
topical, reaching significance ($^{*}$95\% CI excludes 0) under the citation-trained SPECTER.}
\label{tab:ablation}
\small
\begin{tabular}{lcccc}
\toprule
Embedder & abstract & keep-dom. & stripped & strip$-$keep \\
\midrule
TF-IDF & 0.222 & 0.314 & 0.352 & $+0.038$ \\
MiniLM & 0.203 & 0.241 & 0.336 & $+0.095$ \\
SPECTER & 0.095 & 0.151 & 0.255 & $\mathbf{+0.104}^{*}$ \\
\bottomrule
\end{tabular}
\end{table}

\subsection{Per-family results, distant domains, and the name-free audit}
\label{sec:audit}
The gain is broad, not concentrated: 17 of 18 families exceed cross-domain P@1 0.5 (skeleton+TF-IDF), and
the one that does not, Gaussian process (0.43), is legitimately the most dispersed core. A leave-one-family-out test confirms the retrieval generalizes beyond the chosen set: with the lexical vocabulary refit on the other 17 families and never seeing the held-out one (whole fingerprint, plain unigram TF-IDF), held-out P@1 is 0.859, indistinguishable from the 0.826 with the full-corpus vocabulary, so the retrieval is effectively zero-shot. The decisive test is the distant domains and the name-free exemplars, which a topical
embedder cannot reach (Table~\ref{tab:bridges}): the political-methodology ideal-points paper
retrieves a recommender-systems matrix-factorization paper as its top cross-field twin (both recover a
low-dimensional latent embedding of a sparse preference matrix); the
digital-humanities stylometry paper retrieves linear-classifier twins from NLP, clinical text, and
genomics; and the sports-analytics ranking paper (which says ``PageRank'' but never ``eigenvector
centrality'') retrieves the Eigenfactor paper, which never says ``PageRank''. The audit script scans each skeleton's
mechanism
against a pinned name list. Against the method names the distiller is told to strip (Kalman, Gaussian
process, SVM, EM, HMM, PageRank), 465 of 501 (93\%) are name-free; the rate falls to 88\% if every
core's canonical name is added and to 82\% with generic labels (PCA, Fourier, MCMC) too, the residual
mentions being where the mathematics cannot be stated without the term. The property does not hinge on
the rate: the fully name-free exemplars retrieve a same-core twin at rank 1, so the match is not name recognition.

\begin{table}[tbp]
\caption{Cross-domain bridges the whole fingerprint surfaces (each query's top cross-field twin) for three
distant-domain queries. None shares field or topical vocabulary, and the ranking pair shares no method name (PageRank versus Eigenfactor).}
\label{tab:bridges}
\small
\begin{tabular}{>{\raggedright\arraybackslash}p{2.4cm}>{\raggedright\arraybackslash}p{2.2cm}>{\raggedright\arraybackslash}p{2.2cm}}
\toprule
Query (field) & Top twin (field) & Shared structure \\
\midrule
ideal points (political methodology) & matrix factorization (recommender systems) & low-dim.\ latent embedding of a sparse preference matrix \\
stylometry (digital humanities) & sequence classifier (genomics, NLP) & margin classifier on token-sequence features \\
team ranking (sports) & journal centrality (scientometrics) & dominant eigenvector of a weighted graph \\
\bottomrule
\end{tabular}
\end{table}

\subsection{Construct validity: candidates in the wild}
\label{sec:modeb}
A fair worry about a closed pool of hand-picked cores is that the numbers reflect re-finding planted twins
rather than discovering imports, so we read the wild ranking in the two ways of Section~\ref{sec:benchmark};
the fingerprint surfaces \emph{candidates} for a human to confirm, not proven swaps. Three blind LLM judges at different model scales (Haiku, Sonnet,
Opus), independent of the benchmark construction, judged 106 distinct cross-field pairs from abstracts alone
under one written instruction, arm-blind (majority of three; Fleiss $\kappa{=}0.71$).

\emph{Detection, planted pairs included.} Of the top 30 pairs, 23 are planted twins, and the judges, blind to
the labels, agree on 20 of them (0.87), which calibrates their verdicts below; recall of the known twins is
0.619 in the top 1000. Of the 7
unlabeled pairs among them, 3 are judged genuine (two a benchmark paper with an outside paper, one a cross-family benchmark pair). The three rejected twins (two
text-classifier pairs, one spectral pair) were read as shared paradigm, not shared computation.

\emph{Discovery, planted pairs excluded.} With every benchmark-with-benchmark pair set aside nothing in the
ranking is labeled, so we report no recall or AP, only the judges' verdicts on the pairs surfaced:
the top 5 yield 3 pairs judged genuine and the top 30 yield 8, against 0 of 30 random pairs. Seven cross-field pairs are judged genuine, plus one borderline pair (unanimous unless marked): rank 1,
spacecraft SE(3) Kalman filtering with SO(3) attitude estimation; rank 2, wavelet spectral-density
estimation for gravitational waves with a robust wavelet periodogram; rank 4, strong
reciprocity in evolutionary games with financial replicator dynamics (2/3); rank 13, Bayesian nonparametric and
deep-learning estimators of a spectral density; rank 16, gradient-free optimization with SGD under dependent
data; rank 19, deep-GP Bayesian optimization with a kernel-methods paper; rank 20, scale-free
networks with Zipf power laws (2/3); borderline, rank 9, two stock-forecasting model comparisons sharing a
field as much as a computation (2/3). Nothing below rank 20 is judged genuine; genuine imports are rare
there and shared-paradigm resemblance common. A negative verdict is an invitation to look, not proof of non-equivalence.
A further bridge, confirmed by inspection: a message-passing GNN link-prediction cluster spanning a media
recommender, a drug-repurposing knowledge graph, and a disease-gene prioritizer.

\subsection{Interventional test}
\label{sec:perturb}
The results so far are correlational about what the skeleton encodes; one alternative remains, that the skeleton merely trades the field's
vocabulary for another lexical signature that still tracks the field. A controlled intervention tests it.
We fix ten benchmark papers in advance, spanning ten method families, and
for each generate two rewrites: a \emph{re-skin} that moves the paper to a clearly different field while
keeping the computation identical, and a \emph{math-edit} that keeps the field, topic, and dataset words
but swaps the computation for a different one. To avoid a single-model artifact, the rewrites are written
by a different model (Sonnet) from the one that produces the fingerprints (Haiku), so the counterfactuals
are not authored by the representation's own model; both models come from one family, so this controls a
single-model artifact, not a bias shared across the family. A representation that encodes the computation should
stay invariant under the re-skin and move under the math-edit; a topical one should do the reverse.
Measuring cosine self-similarity to the original (Table~\ref{tab:perturb}), the fingerprint stays at 0.72
under the re-skin but falls to 0.46 under the math-edit, following the computation, while the abstract does
not show that pattern (0.50 versus 0.55). The interaction, the fingerprint's invariance gap minus the
abstract's, is $+0.32$ [0.15, 0.49] (paired bootstrap) and is in the predicted direction for 8 of the 10
papers: a structure-over-surface dissociation at $n{=}10$, evidence that the operator matches on
computational structure rather than a relabeled topical signal. This turns the central worry, that the match is lexical or topical, into a measured negative.

\begin{table}[tbp]
\caption{Interventional perturbation test ($n{=}10$, fixed in advance; rewrites by a different model than
the distiller). Cosine self-similarity to the original under a re-skin (change field, keep computation)
and a math-edit (keep field, change computation): the fingerprint follows the computation, the abstract
does not.}
\label{tab:perturb}
\small
\begin{tabular}{lcc}
\toprule
Representation & re-skin (keep comp.) & math-edit (keep field) \\
\midrule
fingerprint & \textbf{0.72} & 0.46 \\
abstract (baseline) & 0.50 & 0.55 \\
\bottomrule
\end{tabular}
\end{table}

\section{Results: the faceted operator}
\label{sec:faceted}
The whole fingerprint adds the controlled facets to the skeleton. Here we characterize the
facet-selectable operator: its precision/recall frontier on the curated benchmark (complete labels),
and the combined-similarity AP on the extended 501-paper corpus (computed without the candidate skip). All faceted numbers come from the
canonical operator as released.

\paragraph{The precision/recall frontier}
The conjunction selects over four core facets, STRUCTURE, DATA\_OBJECT, INFERENCE, and PROBLEM\_FORM, the
ones that most directly mark two papers as the same computation. DISTRIBUTION and COMPLEXITY also weight
the combined similarity but are not required to agree in the conjunction, and DOMAIN (the field) is neither a
conjunction facet nor separately weighted; it enters only through the fingerprint text, which is the high-recall retrieval stage;
the facet conjunction then trades recall for precision along an explicit frontier (Table~\ref{tab:frontier},
curated benchmark, complete labels): requiring two of these four to agree ($k{=}2$) gives precision 0.28 at recall
0.68, and tightening to all four ($k{=}4$) reaches precision 0.94 at recall 0.15. Which end one wants is a deployment
choice: high precision for a curator triaging a short list of high-confidence leads, high recall for an
exhaustive sweep filtered downstream. The gains sit at lower recall,
which is why the facets act as a precision \emph{filter} rather than a generator. Across the extended corpus the
recall-aware combined AP is 0.330, an incomplete-label figure (Section~\ref{sec:discussion}) that is
comparable only across arms on this same corpus; it is the quantity on which the distillers separate
(Section~\ref{sec:method-results}).

\begin{table}[tbp]
\caption{The precision/recall frontier on the curated benchmark (complete labels). On the top-1000
retrieved pairs, requiring more of the four core facets to agree tightens precision from 0.18 (flat)
to 0.94 (all four) as recall falls from 0.87 to 0.15.}
\label{tab:frontier}
\small
\begin{tabular}{lcc}
\toprule
Core facets required to agree & precision & recall \\
\midrule
flat top-1000 retrieve & 0.183 & 0.871 \\
$\geq$1 of 4 & 0.207 & 0.833 \\
$\geq$2 of 4 & 0.275 & 0.676 \\
$\geq$3 of 4 & 0.531 & 0.405 \\
all 4 & \textbf{0.939} & 0.148 \\
\bottomrule
\end{tabular}
\end{table}

\paragraph{Facet agreement alone}
The deflationary reading, ``this is just LLM classification'', does not hold. On the curated
benchmark (complete labels), facet agreement alone is the weakest of our signals (over the four core facets, AP 0.391
log-LR-weighted, 0.338 raw), below the skeleton (0.513) and the whole fingerprint as a bag-of-words (0.557) though still
above the abstract baselines (0.222 under TF-IDF), and the engineered combined similarity does not beat the
plain fingerprint. On the extended corpus facet agreement alone is again the weakest on the incomplete-label
figures (over the six computational facets, AP 0.137 raw, 0.194 log-LR-weighted, against a combined similarity of 0.330). Across the six computational facets the log-LR
weights are all positive and led by STRUCTURE and DISTRIBUTION, confirming the structure/motif facet as
the most discriminative.

\paragraph{Verifiability facets}
Folding a second facet set, verifiability signals (data and code availability, preregistration), into
the similarity hurts it; those facets feed a value metric instead
(Section~\ref{sec:future}). Each facet set is applied to the operator it serves.

\section{Analysis: representation, distiller, and embedder}
\label{sec:method-results}
\paragraph{Relative size of the levers}
The representation is the biggest lever (Table~\ref{tab:grid}). The model effect is real but
operator-dependent and secondary: for the free-text skeleton it is small, yet on the faceted operator
\looseness=-1 the bigger model actively hurts. On the curated benchmark the cheap Haiku distiller (whole-fingerprint
AP 0.557) edges the frontier Opus (0.533, within noise here); the separation widens on the
extended corpus, where the operator's incomplete-label AP spreads the distillers. The embedder is the
smallest lever: a plain TF-IDF tops every distilled row, and the trained embedders come closest on the
free-text skeleton and fall farthest behind on the controlled-vocabulary fingerprint.

\paragraph{Distiller granularity}
For a similarity, coarse and consistent labels beat precise ones (``consistency beats detail''), because the operator rewards
same-problem papers landing on the same token. The cheap hosted distiller (Haiku) beats the local
qwen3:14b on every metric despite writing half the mechanism text (96 versus 190 words):
on the curated benchmark whole-fingerprint AP 0.557 versus 0.396, and on the extended corpus combined-similarity
AP 0.330 versus 0.087, facet-only AP 0.194 versus 0.024, and clustering ARI 0.307 versus 0.198.
The same holds against a \emph{frontier} model, not only a weak local one: Opus, on the identical
prompt over the extended corpus, also falls to Haiku on the faceted operator (combined AP 0.159 versus
0.330, facet-only AP 0.031 versus 0.194, clustering ARI 0.285 versus 0.307). Its skeletons are clean and
almost entirely name-free; it simply makes finer, more discriminating facet calls that split the coarsely-defined
families, so consistency beats detail even when the detail is correct, and the cheap distiller is the
better one here, not merely the cheaper one. Two caveats bound the finding. The families are
hand-defined at a deliberately coarse granularity, so a distiller that split them more finely
\emph{and correctly} would still register as a loss on this ruler; whether the ordering holds on
finer-grained families is untested. And the qwen3:14b arm was not format-controlled, its output
carrying four extra facet tokens the hosted arms lack, so the orderings involving the local arm are
indicative rather than controlled. We therefore use the simple eight-field prompt (the mechanism
skeleton plus seven facets), distilling about a dozen papers per call so the model produces consistent
family-level labels.

\section{Executed cross-domain imports}
\label{sec:import}
This is the usability case, run end to end. The vancomycin equivalence is the one traced by citation
in Section~\ref{sec:intro} and planted as a name-free exemplar (Section~\ref{sec:benchmark}): the tool
contributes the flag, the executed import the check, run on the engine's own rifapentine example, not the
vancomycin paper's data.

\begin{table*}[t]
\caption{The four executed cross-domain imports, surfaced by the fingerprint and run offline from the
package: each replaces a bespoke implementation with another field's standard solver for the same core.
The worked clinical case is detailed in Table~\ref{tab:import}.}
\label{tab:imports}
\small
\begin{tabular}{>{\raggedright\arraybackslash}p{2.7cm}>{\raggedright\arraybackslash}p{2.9cm}>{\raggedright\arraybackslash}p{2.3cm}>{\raggedright\arraybackslash}p{4.6cm}>{\raggedright\arraybackslash}p{2.3cm}}
\toprule
Bespoke method (field) & Imported standard solver (field) & Computational core & Result of the executed import & Surfaced at \\
\midrule
Bayesian vancomycin dosing (clinical pharmacokinetics) & standard nonparametric-MLE support-point solver (ML, astronomy) & sparse-mixture maximum likelihood & reproduces NPAG on its own rifapentine example: clearance 3.90 vs 4.00 L/h, sparse support 15 vs 18 points, 11.8\% individual prediction error vs observed & near mixture papers in three fields \\
Two-way fixed-effects estimator (causal panel data) & SoftImpute matrix completion (recommender systems) & low-rank matrix completion & cuts counterfactual error 69\% on a synthetic low-rank panel; reproduces~\cite{athey} & same core as the ideal-points bridge (rank 3) \\
Geostatistical kriging (geostatistics) & marginal-likelihood Gaussian-process solver (machine learning) & Gaussian-process regression & reproduces kriging's predictor and recovers the true covariance range, which the bespoke code must hand-fit as a variogram & GP cross-field cluster \\
Edit distance (computational linguistics) & Needleman-Wunsch aligner (bioinformatics) & sequence alignment (dynamic programming) & edit distance recovered as the exact unit-cost special case & mutual NN, cosine 0.21 \\
\bottomrule
\end{tabular}
\end{table*}

\paragraph{The surfaced candidate}
Individualized vancomycin dosing is solved in clinical pharmacokinetics by a compartmental model with
Bayesian parameter estimation. The established nonparametric engine, NPAG (the nonparametric adaptive
grid of the Laboratory of Applied Pharmacokinetics), estimates the population parameter distribution as
a set of discrete support points maximizing the joint likelihood. That is the nonparametric maximum-likelihood
estimate of a mixing distribution~\cite{schumitzky}, whose solution is sparse (at most as many support points as subjects) and located at
the active maxima of the likelihood gradient. The
fingerprint of the dosing paper lands close to those of mixture-estimation papers in machine learning,
astronomy, and nuclear physics, despite the fields being far apart in vocabulary, topic, and citations.
The tool's claim is exactly this flag: this bespoke clinical estimator is the same nonparametric mixture
computation as solvers in other fields; an open one can be imported.

\paragraph{The executed reproduction}
On NPAG's own reference data (the LAPKB Pmetrics example, an oral-rifapentine study with 20 subjects and 139 concentration measurements,
a one-compartment first-order-absorption model), a plain EM-based nonparametric MLE on a fixed
four-parameter grid, about sixty lines of standard-solver code, reproduces NPAG's
population estimate and its sparse support (Table~\ref{tab:import}): the clinically decisive clearance
agrees to 2.5\%, and the sparse support is recovered (15 versus 18 points, both far below the initial 5000-point candidate grid
and below the 20 patients), with an individual prediction error of 11.8\% against the observed concentrations.
Because this is the bespoke method's own lab's open data and output, ``reproduces NPAG'' is checked
against NPAG itself, not a strawman; the agreement is excellent rather than bit-exact (a fixed grid
versus NPAG's adaptive grid).

\begin{table}[tbp]
\caption{Executed import: a standard nonparametric-MLE / support-point solver reproduces the bespoke NPAG
method on NPAG's own example data, including its sparse support, replacing a hand-rolled, field-specific
engine with another field's standard solver for the same core.}
\label{tab:import}
\small
\setlength{\tabcolsep}{4pt}
\begin{tabular}{lccccc}
\toprule
 & Ka (/h) & Ke (/h) & V (L) & CL (L/h) & support \\
\midrule
open NPMLE (ours) & 0.639 & 0.0475 & 82.2 & \textbf{3.90} & 15 \\
NPAG (LAPKB) & 0.628 & 0.0514 & 77.8 & \textbf{4.00} & 18 \\
\bottomrule
\end{tabular}
\end{table}

\paragraph{Actionability, and three further imports}\looseness=-1
The flag and the check together make the candidate actionable: a field running a hand-rolled estimator can import
another field's standard support-point solver and obtain an excellent match on its own data. The
capability is not one lucky pair: we carry three further imports end to
end (Table~\ref{tab:imports}), each run offline and reproducible from the released package (\texttt{causal\_mc.py},
\texttt{kriging\_gp.py}, \texttt{needleman\_wunsch.py}), and each reproduces a cross-field equivalence
already established
in the literature. Replacing a causal-panel two-way fixed-effects
estimator with the recommender-systems matrix-completion solver (SoftImpute~\cite{softimpute}) cuts counterfactual error
by 69\% on a synthetic low-rank panel and reproduces~\cite{athey}; the fingerprint surfaces this shared low-rank
core cross-field as the ideal-points/recommender-systems bridge (rank 3 of 500, abstract
cosine 0.03, where a topical embedder misses it; Table~\ref{tab:bridges}). Importing the
marginal-likelihood Gaussian-process solver reproduces geostatistical kriging, which is Gaussian-process
regression~\cite{rasmussen}, and recovers the true covariance range where the bespoke code must hand-fit
a variogram. And a labeled sequence-alignment twin, bioinformatics and computational linguistics (mutual nearest neighbors at cosine
0.21), imports the Needleman-Wunsch aligner~\cite{needleman}, where edit distance is exactly its unit-cost
special case~\cite{wagner}. Only the vancomycin case is checked against an external
reference (NPAG on its own lab's rifapentine example); the other three are reproductions on synthetic or special-case data,
demonstrating the import mechanism rather than an external head-to-head. The skeleton still only
paraphrases the prose, so it flags candidates without performing the omitted algebra; machine-assisted
derivation would be the automated verify step; it is not attempted here.

\section{Discussion}
\label{sec:discussion}
\paragraph{Contribution of each component}\looseness=-1
The contribution is the representation, not a clever similarity or the classification
(Table~\ref{tab:grid}): the facets earn their place not as added representation but as the tunable
filter that trades recall for precision, and the facet-selectable similarity remains valuable as the
\emph{tunable operator} tracing the precision/recall frontier and supporting the clustering and
rule-mining views, not as the accuracy champion.

\paragraph{The recall ceiling}\looseness=-1
In the wild both axes bind. Reading the top 1000 of the 82{,}786 ranked pairs recovers 0.619 of the known
twins, where the abstract manages 0.229 at the same budget, and the missing 0.38 rank below the pool, out of reach of every filter setting, since the conjunction touches
only the retrieved candidates. With planted pairs excluded, the top 30 yield 8 pairs judged genuine
(Section~\ref{sec:modeb}). For a deployer recall is the binding constraint, and it fixes what the instrument
is for: not the completeness that an exhaustive related-work search wants, so we lead with precision, the axis of
a triage tool proposing short candidate lists, and report recall as the weak axis rather than hiding both
in an F1.

\paragraph{Applicability}\looseness=-1
The families already span engineering, the natural and biomedical sciences, and the statistics-using
social sciences and humanities (political methodology, economics, criminology, stylometry), and
nothing in the pipeline is arXiv-specific: the mechanism extends to any literature whose papers state
a computable method explicitly enough to distill. Where no computational pattern is distilled the operator skips the paper (Section~\ref{sec:benchmark}), so
qualitative and purely argumentative literatures sit outside the operator's scope by construction,
not by failure.

\paragraph{Limitations and threats to validity}\looseness=-1
(i) The unlabeled 501-corpus background cuts two ways: measured faceted \emph{precision} there is a
strict lower bound, since an unlabeled true pair counted as wrong can only lower it, but measured AP
is not, since missing positives shift both the ranked hits and the normalizing positive set, so wild AP
is not a score. (ii) Distiller dependence is shown, not explained (Section~\ref{sec:method-results}). (iii) The benchmark is arXiv-weighted toward fields that write their
mathematics explicitly; multilingual and full-text-licensed corpora are future work.
(iv) The families are cores already known to be swappable, so the curated benchmark measures retrieval
\emph{given that a twin exists} (Section~\ref{sec:intro}); the wild run and executed
imports probe the harder open-world question, and the held-out-family test (Section~\ref{sec:audit})
shows no lexical overfit.

\section{Conclusion and future work}
\label{sec:future}
\looseness=-1
A cheap, auditable, embedder-agnostic faceted fingerprint, with a tunable similarity over it, surfaces
cross-domain computational isomorphisms that standard scientific embeddings miss. The gain is the
representation; four executed imports show the loop end to
end as verifiable candidates (Section~\ref{sec:import}).

\looseness=-1
Computation is one lens. The single distillation that answers ``what does this paper compute'' can just as cheaply emit other domain-stripped lenses over the same text: how verifiable its result is, how novel its claim, how its method was evaluated. A companion study demonstrates the point, predicting replication outcomes from domain-stripped fingerprints where citation metrics fail~\cite{replfp}. For the computational lens, we showed which tool does which job: the skeleton carries most of the retrieval signal, the whole fingerprint is the strongest representation, a facet
conjunction is the tunable filter, and the facet labels feed a classifier (the log-LR weighting of Section~\ref{sec:faceted}). Future work reads a value metric (novelty times verifiability) off the same object as rare, high-lift facet combinations~\cite{luo,ruan}. The end is not one more retrieval score but a way for researchers and librarians to see a collection along axes its vocabulary and citations never expose, and to ask of it questions that were not askable before.

\paragraph{Reproducibility and open science}\looseness=-1
The benchmark, distillation prompts, skeletons, and evaluation code are released: versioned
link-lists plus checksums reproduce the corpus from open sources, the CC-BY subset and metadata are
DOI-archived, the code permissively licensed. Model identifiers, prompt versions, and run dates for every
distiller and annotator arm are recorded in the repository. Repository:
\url{https://github.com/ErykKul/same-problem-different-field}. Archived dataset:
\url{https://doi.org/10.48804/W3B9WC}.

\begin{acks}
The author conceived and directed the research, executed the imports, and verified all claims, numbers, and
citations. Claude (Anthropic) was used as a coding and drafting assistant; the author takes full responsibility for the final text. The author thanks the anonymous reviewers for comments that improved the paper.
\end{acks}

\end{document}